Foresight AND Hindsight. Heinz von Foerster's "The cause lies in the future."

Mihai Nadin, Director, antÉ-Institute for Research in Anticipatory Systems, The University of Texas at Dallas, USA

Abstract: To model is to represent. The threshold of decidability defines two epistemological choices: one model (or a finite number of models) suffices for representing the dynamics below the undecidable; above this threshold (defined as G-complexity), every model is partial, no complete modeling is possible.

## Questions asked, questions not answered, answers raising new questions

The only reason for engaging in the Open Peer Commentary of the text by Hannes Hornischer et al. (2020) is the multitude of questions it invites. These go beyond Heinz von Foerster's *Ethical Imperative* and its computational interpretation instantiated as a Future State Maximization machine. All compliments (even the best-intended) are poisonous. The multitude of questions (the compliment) comes loaded: Why were they not answered? And if answered, how many answers raise more questions? This might qualify as a new imperative: always increase the number of questions to the benefit of more interaction. Of course, this assumes that dialog is desired and not only words of praise are allowed.

In the absence of a convincing case for distinguishing between simulation and model, the Introduction argues that although "there seems to be an even more fundamental and generic principle under which model use can be subsumed" (p. 36),  the way to go remains through models. The more "fundamental principle" turns out to be another model. If for no other reason because (so the argument) "space and time are constructions emerging from the use of models." For those conversant in the subject of space and time, especially from a constructivist perspective, the statement begs for arguments beyond examples (as the authors provide).

Long before the authors chose bacteria behavior, Lynn Margulis (1995), and later John Sowa (2016) and Mihai Nadin (2018) dealt with exactly the same. Goal-directed actions identify intentionality. But there is more to this subject, and one can only commend the authors for providing the opportunity to make the point. Change in respect to entities defined as living is expressed through their dynamics in the dynamic environment. The Petri dish is a reduction for making a point, but space and time do not emerge in the Petri dish, and not even in the larger framework (p. 37) in which a "gradient of higher concentrations" is hypothesized. Actually, to be alive is to be active. Goethe, who was also a scientist (some prefer the qualifier *Naturphilosoph*, i.e.philosopher of nature)  ascertained that we know ourselves through action, a thought that Nikolai A. Bernstein, the visionary researcher of motoric expression and implicitly of anticipatory expression (1947) expressed as (I am paraphrasing), *We know the world through our actions*. Take note of the fact that this is no longer Pavlov's understanding of reflex, i.e., reaction to the world, but "activity means purposeful action" (1936).

In the discussion ("short detour"), modeling gets stuck in circular thinking: How should the virtual reconstruction of what is at stake in the given situation—bacteria able to somehow procure nutrient—take place? According to von Foerster, in the absence of experience, the famous corollary "the environment contains no information; it is as it is" (1974)—the basic constructionist epistemology—stands. Before the constructivist, and in sharp contradistinction to Pavlov, Bernstein took note that to move in the world is to know it, not virtually, but existentially, or through acquired experience. The slime mold (Ball 2008) and also plants and everything else that is alive behave in a manner informed by experience. The specific movements of archaea, fungi, protists, bacteria, plants, and animals are the expression of reactiveness—what further down the line will be described as the outcome of hindsight—AND activeness—i.e., foresight. The former is by its nature deterministic, the latter, non-deterministic (Nadin 2020)[1]. Of course, this distinction invites some details. Foresight (Προμηθεύς) and hindsight (Ἐπιμηθεύς) are concepts reflecting the pragmatic framework of the context in which they are defined. It is more a matter of shared understanding.

Since we are what we do (Nadin 1997), we constantly construct and reconstruct ourselves and implicitly our concepts. In the contemporary context of justified interest in computational representations, the two concepts are by necessity reinterpreted in terms pertinent to what computation is. In von Foerster's words: "Experience is the cause, the world is the consequence." In 1995, the American Society of Cybernetics made public an *Anthology of Principles*…[2] on the subject of circularity containing the above sentence.

---

[1] Nadin M. (2020a) Reactia la crize: Prometu si Epimeteu. Un epilog. https://www.revistasinteza.ro/mihai-nadin-reactia-la-crize-prometeu-si-epimeteu-un-epilog. Sinteza: a journal for culture and strategic thinking. July 18. Cluj, Romania. (The English text is available at https://philarchive.org/rec/NADPAE?all_versions=1

[2] Foerster H. von (1995) Anthology of principles propositions theorems roadsigns definitions postulates aphorisms etc. H.V.F. Cybernetics and Circularity. May 17-21. http://www.cybsoc.org/heinz.htm

Deriving his ethical Imperative[3] under Immanuel Kant's categorical imperative example, von Foerster constructed, implicitly, the notions of action, possibility, and choice. Other tried a different path. To paraphrase Ernst von Glasersfeld (1984): We face reality as a burglar in the castle. In order to get to the booty, the burglar has to unlock it. The bacterium in the article—as well as all of us in the time of SARS-CoV-2—are in the same position. There is no multiplication of choices, rather an obsession with finding the key. In our days, *code* is the concept used.

It is highly commendable that Hornischer et al. are looking for a principle different from that of traditional AI and machine learning. The originators of this path—Henry J. Charlesworth and Matthew S. Turner (2019) examine the emergence of social cooperation in animals. Working in groups give members advantages. The model of the swarm (or the flock, as with birds) proved to be effective for particular applications (many of them military in nature). In short, simulated birds navigate a virtual world (images). Maximizing the space of options results in cooperative behavior—it is not pre-programmed. This particular technique goes back to the time when Artificial Life (ALife) was fashionable (and generously funded). Upon a closer look, Future State Maximization (FSM; but since the authors opted for FSX, their option is adopted *volens-nolens* in this commentary) recognized a valid question: Birds or not, is this an organizational principle characteristic of what is usually (i.e., at a primitive level ) described as intelligence? They suggest it is. How intelligence is defined remains an open question. Hornischer et al. called up a

---

[3] Foerster H. von (1995) Anthology of principles propositions theorems roadsigns definitions postulates aphorisms etc. H.V.F. Cybernetics and Circularity. May 17-21. http://www.cybsoc.org/heinz.htm

Workshop (December 2019, Graz, at their University). One of the contributions (Margarete Boos and Hannes Hornischer 2019[4]) was on maximizing options—an underlying principle of collective behavior in human groups. But neither in the article subject to this Commentary, nor in the mentioned workshop presentation is the above question answered.

## If computation, then what kind of computation?

In fact, the birds in the computational experiment are as much birds as virtual space is space. What is maximized is a computer-generated space representation—pixels, or vectors, or whatever it takes to translate space geometry into virtual geometry. Once again, those who took time to understand FSX had no difficulty in realizing that the algorithm (tree searches, nothing else) is similar to the algorithm behind any known game machine (chess, Go, but also machines playing computer and video games). Consider possible game options and select (using some weighting procedure) those "moves" that promise the highest return. The principle, which remains indeed in the modeling technique repertory, is applied in applications such as robotics or even digitally assisted farming, also known under the misleading title Precision Agriculture. Based on weather predictions (too dry, to wet, etc.),  insurance companies advise a farmer whether to raise a crop or cash in for not cultivating the land for a season. Cooperative behavior of agents used in such applications is algorithmic. Keeping options open maximizes the level of cooperation. If you use Excel or program in Python you know that the *max*() function returns the item with the highest value, or the item with the highest value in an

---

iterable. That's all there is. No intelligence, since there is no understanding of the meaning of the data processed.

In view of the above, the discussion of the *Ethical Imperative* appears as an attempt to align it with FSX. The Maturana and Varela (1987) metaphor, which the authors mention in order to suggest the relation between maximizing future options and operational closure, disclose the reactive choice: sensors provide the submarine pilot with enough information for successful navigation in a dangerous deep-sea landscape. It is a deterministic machine, without any pro-active capabilities. In other words: no foresight at work.

This puts their own rhetorical question—"Foresight rather than Hindsight?"—in a context beyond the rhetoric. Indeed, when "computable" means *only* algorithmic computation—in the sense of a Turing machine—and when information is actually *only* data—as in Shannon's model, which is strictly syntactic (no semantic dimension, as Shannon himself made clear)— what results is a deterministic machine. Regardless of the way it is implemented, it can *only* represent deterministic processes. The examples (agent positioned on a 2-dimensional finite grid, §19-24), emergence of leadership in groups (Example 1), aquatic robotic swarm, (Example 2 ) deserve respect for generating machines that operate according to the FSX. More problematic is the "Learning to behave" (§44). The authors claim that "the agent's model emerges as information about probability of being in a certain state after taking a particular action in an initial state." The pseudo-code does not justify the sentence "the model emerges as an eigenvalue." The number of states sampled (random choices) and the learned model are in a predefined relation; therefore "eigenvalue" (described as "iterated attempts of an agent to come to terms with its options") as a kind of attractor is simply not an option. Demon dialers assisted by neural networks do better, without any claim to learning. I am inclined to identify here the misunderstandings of a text by Füllsack and  Riegler A. (2017) on "thinking in eigenbehaviours"—ideas can get easily contaminated by questionable references.

Be this as it may, the process described by the authors,  and the description are congruent. This quality cannot be noticed when the notion of anticipation is brought up (§53). To "anticipate the most sustainable states, i.e., the one that provides highest resilience, under conditions of non-linearity and sudden, unpredictable transitions" (examples would be climate change) would imply a daemon, or at least a different kind of computation (non-algorithmic). In short, the authors are not really willing to invest the effort in understanding the nature of anticipatory processes.

## The Ursache liegt in der Zukunft

Discussing questions that the paper inspires is, of course, different from observing that prediction—i.e., to ascertain something about the future—is quite different from anticipatory action—i.e., to be active in the process of change. Prediction reflects the degree to which a deterministic process is captured in an actionable mathematical description. Anticipatory action, based on learning (experience remains the backbone of a constructive view of the world) is non-deterministic. The bacterium described in the paper actively discovers the world; a food could sustain life, or, if it is, by accident or by design (through a different subject) poisonous (bait), it can end life, or handicap it. In none of the examples given in the article can anticipation action emerge, because none of the computational processes involved has agency.

There is a past from which data (not information) are derived, and there is a syntactic pattern. Anticipation ensues on account of learning, it undergirds evolution, and is manifested as consequential foresight. Ethics in von Foerster's view, is anticipatory. Pursuant to a long conversation with him (Pescadero, CA, May 1999) on a formulation that invited unpacking its meaning—"Die Ursache liegt in der Zukunft"[5]— I adopted the title (with his blessing) for my book (Nadin 2003) *Anticipation—The End is Where We Start From* (actually quoting T.S. Eliot). But this goes, at least in this succinct formulation, beyond the limits of a commentary.

More relevant is the understanding of the nature of von Foerster's Ethical Imperative. Gödel's undecidability offers a suggestive clue. It turns out that the living is G-complex (Nadin 2014), i.e., does not allow for decidable representations (i.e., descriptions or models).

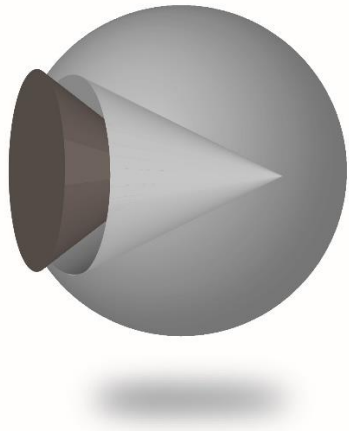

Figure 1. The undecidable and the possibility to select a certain aspect of an undecidable process in order to achieve, for the subset, both completeness and consistency

If indeed the increase in the number of choices should inform human action, the space of choices and the time during which they are made are open-ended. In short, we can never describe them all (they continue to multiply), and they will not be consistent. Algorithmic computation is decidable by definition. The article succinctly names its results: FSX corresponds to the Ethical Imperative. No proof for this was produced—because it cannot be produced. The world is not flat.

To express gratitude to the authors for offering a lot of material to think about and debate is my sincere compliment. No flattery intended.

## Bio

Nadin's professional life combines engineering, mathematics, digital technology, philosophy, semiotics, mind theory, and anticipatory systems. He holds advanced degrees in Electrical Engineering and Computer Science, and a post-doctoral degree in Philosophy, Logic and the Theory of Science. With his book, *To Live Art* (1972), Nadin was a precursor of the recent Anticipatory Systems research. While teaching in Europe, Nadin established the Institute for Research in Anticipatory Systems (2002), which became part of the University of Texas at Dallas when he accepted the invitation to become Ashbel Smith University Professor. Granted a fellowship at the prestigious Hanse Institute for Advanced Study (Germany), Nadin researched (since 2011) anticipatory systems within the neuroscience group, and established the international study group on anticipation across disciplines. For more information and full access to his publications, see www.nadin.ws